# Residual-based physics-informed transfer learning: A hybrid method for accelerating long-term CFD simulations via deep learning

Joongoo Jeon[a,b,1], Juhyeong Lee[c,1], Ricardo Vinuesa[a] and Sung Joong Kim[c,d*]

jgjeon41@jbnu.ac.kr, juhyeonglee@hanyang.ac.kr, rvinuesa@mech.kth.se, sungjkim@hanyang.ac.kr

[a]FLOW, Engineering Mechanics, KTH Royal Institute of Technology

SE-100 44 Stockholm, Sweden

[b]Graduate School of Integrated Energy-AI, Jeonbuk National University

567 Baekje-daero, Deokjin-gu, Jeonju 54896, Republic of Korea

[c]Department of Nuclear Engineering, Hanyang University

222 Wangsimni-ro, Seongdong-gu, Seoul 04763, Republic of Korea

[d]Institute of Nano Science & Technology, Hanyang University

222 Wangsimni-ro, Seongdong-gu, Seoul 04763, Republic of Korea

[1]These authors equally contributed to this work.

[*]Corresponding author: Email: sungjkim@hanyang.ac.kr (Prof. Sung Joong Kim)

**Abstract**

While a big wave of artificial intelligence (AI) has propagated to the field of computational fluid dynamics (CFD) acceleration studies, recent research has highlighted that the development of AI techniques that reconciles the following goals remains our primary task: (1) accurate prediction of unseen (future) time series in long-term CFD simulations (2) enhanced caclulation performance (3) an acceptable amount of training data and time (4) within a multiple PDEs condition. In this study, we propose a residual-based physics-informed transfer learning (RePIT) strategy to achieve these four objectives using ML-CFD hybrid computation. Our hypothesis is that long-term CFD simulation is feasible with the hybrid method where CFD and AI alternately calculate time series while monitoring the first principle's residuals. The feasibility of RePIT strategy was verified through a CFD case study on natural convection. In a single training approach, a residual scale change occurred around 100th timestep, resulting in predicted time series exhibiting non-physical pattern as well as a significant deviation the ground truth. Conversely, RePIT strategy maintained the residuals within the defined range and demonstrated good accuracy throughout the entire simulation period. The maximum error from the ground truth was below 0.4 K for temperature and 0.024 m/s for x-axis velocity. Furthermore, the average time for 1 timestep by the ML-GPU and CFD-CPU calculations was 0.171 s and 0.015 s, respectively. Including the parameter-updating time, the simulation was accelerated by a factor of 1.9. In conclusion, our RePIT strategy is a promising technique to reduce the cost of CFD simulations in industry. However, more vigorous optimization and improvement studies are still necessary.

**Keywords:** physics-informed machine learning; computational fluid dynamics; residuals in governing equations; residual-based transfer learning; simulation acceleration

# 1. Introduction

In the early 1800s, Stokes from England and Navier from France derived nonlinear partial differential equations describing the relationships between the velocity, pressure, temperature, and density of a moving fluid [1]. These remarkable equations called the Navier–Stokes equations can solve real-world viscous flow problems by adding a viscous effect term to the Euler equation. Under general flow conditions, analytically solving the Navier–Stokes equations is difficult. In the past, many studies aimed to solve these equations by simplifying them through several assumptions and approximations [2]. However, advances in computer performance have allowed solving them numerically, instead of analytically which limit the solvable flow conditions. Various numerical schemes such as finite difference [3], finite volume [4], finite element, and spectral methods [5] have been adopted to investigate fluid behavior including heat transfer and are called computational fluid dynamics (CFD) simulations. Advances in CFD have contributed to various fields such as nanofluids [6-8], weld formation [9, 10] and energy safety [11, 12].

However, despite the rapid advancements in the performance of central processing units (CPUs), the computational cost of simulating unsteady flows having extremely small time/grid-scale physics is still unrealistic. These complex flows are typically turbulent flows possessing a time-averaged Reynolds stress tensor or reacting flows requiring thermal radiation and reaction heat calculations. Jeon et al. observed that simulating a turbulent jet under pipe rupture conditions required 70 h of CPU time per second of physical time (CPU time/s), with parallelization across 32 cores [4, 11]. Recently, laminar flame simulations required 16 h CPU time/s with parallelization across 32 cores [12, 13]. Tolias et al. needed approximately 10–100 h CPU time/s (with parallelization across 4–32 cores) to investigate unsteady hydrogen deflagration by CFD simulations based on a domain size of 20.0 × 14.4 × 12.0 m [14]. Even with parallel core computation, simulations on the scale of minutes and hours are extremely limited on general laboratory unit machines.

In recent years, machine learning (ML) technology has received the most attention across industries, and this major trend has generated various interests in the fluid dynamics community [15, 16]. A summary of the studies applying ML techniques to CFD simulations can be found in Ref. [17]. As Vinuesa noted, one of the main motivations of applying ML is the acceleration of conventional CFD [18]. Early studies have demonstrated that networks trained using past time series data can predict near future time series with good agreement [17]. However, these studies have also revealed that completely suppressing the error increase caused by the increase in the interval between the training and prediction times is unrealistic [17-19]. In conventional data-driven methods, simulating long-term unseen time series is virtually impossible. We outline recent mainstream neural network-based approaches for addressing the issue of unseen data in partial differential equations (PDEs).

- Convolutional neural networks (CNNs): The success of CNNs in computer vision has demonstrated the usefulness of the CNN architecture, especially for image processing. Because CFD snapshots can also be considered as image data, CNN-based approaches have recently been attempted to predict CFD variable fields. Gou et al. demonstrated that steady flow fields under different obstacle conditions can be predicted using a single training set of 100,000 field images [20]. However, a greater number of images is needed for network training for unsteady flow prediction. Recently, Lee et al. identified that a network trained on 500,000 images can predict the unsteady flow fields over a circular cylinder using a multi-scale CNN method [19]. They confirmed that the multi-scale CNN, considering conservation laws, was shown to be more predictive than conventional approaches, but the mechanism for production of small-scale vortical structures became increasingly difficult to predict over time [19].

- Physics-informed neural networks (PINNs): Recently, Raissi et al. developed PINNs which is one of the most innovative neural network-based approaches for PDEs calculation [21]. The principle of PINNs can be understood through the algorithm below. Because the loss function is constructed by the PDE equations, PINNs is a type of unsupervised learning. One more attractive feature of PINNs is that it is a meshless solver because the PDE equation is directly computed using automatic differentiation. Despite their ability to predict unseen time series, the disadvantage is that the calculation speed is slower than CFD solvers [22]. Go et al. identified that the heat conduction problem which took 35 s with a CFD solver, took 34,035 s to learn with a PINN-based model. They concluded that it is inefficient to use PINNs as a standalone solver rather than as a surrogate model [22].

| The PINN algorithm for solving differential equations [23] | |
|---|---|
| Step 1 | Construct a neural network $\hat{u}(\mathbf{x}; \boldsymbol{\theta})$ with parameters $\boldsymbol{\theta}$. |
| Step 2 | Specify the two training sets $\mathcal{T}_f$ and $\mathcal{T}_b$ for the equation and boundary/initial conditions |
| Step 3 | Specify a loss function by summing the weighted $L^2$ norm of both the PDE equation and boundary condition residuals |
| Step 4 | Train the neural network to find the best parameters $\boldsymbol{\theta}^*$ by minimizing the loss function $\mathcal{L}(\boldsymbol{\theta}; \mathcal{T})$. |

- Deep operator network (DeepONet): Another idea that can effectively applies neural networks to PDEs calculation is neural operator approaches. Recently, Lu et al. developed DeepONet using deep neural networks based on the universal operator theorem [24]. The universal operator approximation theorem states that a neural network with a single hidden layer can accurately approximate any nonlinear continuous functional and operator. For example, we can aim to learn the operator that maps the initial condition to the PDE solution at a specific time. Li et al. noted that the neural operator approach is up to three orders of magnitude faster compared to traditional PDE solvers [25]. However, they did not include CFD data production time and training time in their acceleration performance. In their study, 10,000 training pairs $\{a_j, u_j\}$ were needed to learn the Navier-Stokes equation [25]. We need a method that can accelerate the simulation of unseen time series data including training time whenever PDEs or geometry change.
- Finite volume method networks (FVMN): The motivation behind FVMN is to align the network architecture to finite volume method principles and inccorporate the PDEs in the loss function, similar to PINNs [17]. The most attractive feature of FVMN is that the number of training time series required to predict unseen (future) times series is much smaller than in conventional approaches because it takes the form of grid-based supervised learning. **Fig. 1** shows that a network trained with only 3 time series exhibits good agreement with the ground truth for the next 200 time series. However, the error accumulation eventually caused the ML prediction path to deviate from the ground truth path in a long-term simulation. Unlike first-principles simulations, neural networks cannot explicitly solve the Navier–Stokes equations with internal iterations to reduce the residuals. The nonlinear Navier–Stokes equations can be accurately approximated by combining linear and activation functions only within the close time range used for the network training.

The aforementioned efforts highlight that despite much progress, the development of a practical CFD acceleration method that reconciles the following three goals remain our outstanding challenge: (1) accurate prediction of unseen (future) time series in long-term CFD simulations (2) faster calculation performance than CFD solvers and (3) an acceptable amount of training data/time for each PDEs and geometry change. It should be noted that the methods introduced above mainly investigated their accuracy and acceleration performance under a single PDE condition. Therefore, this study aims to (4) verify the performance of the new method under a multiple PDEs condition.

Our hypothesis is that long-term CFD simulation is possible with a hybrid method where CFD and AI alternately calculate time series throughout the entire timeline. In our hybrid method, the role of the CFD solver is not only to reduce the increased residuals but also to update the network parameters with the latest CFD time series data (transfer learning manner). Additionally, if the cross point of ML↔CFD can be pinpointed based on the first principle's residuals (physics-informed approach), the ground truth of future time series is unnecessary for our hybrid method (unsupervised manner). Therefore, the objectives of this study are (1) to develop a hybrid method that can meet the previous four goals, based on the physics-informed transfer learning approach and (2) to validate the accuracy and acceleration performance of this method using an unsteady CFD dataset. To the best of the authors' knowledge, this is the first study to apply the ML-CFD hybrid computation method in the form of transfer learning to accelerate heat and mass transfer simulations. *Section 2* introduces the novel ML strategy; residual-based physics-informed transfer learning (RePIT) strategy. *Section 3* describes the CFD dataset. *Section*

*4* compares the performance of this strategy with that of a single-training approach. Finally, *Section 5* summarizes and concludes this paper.

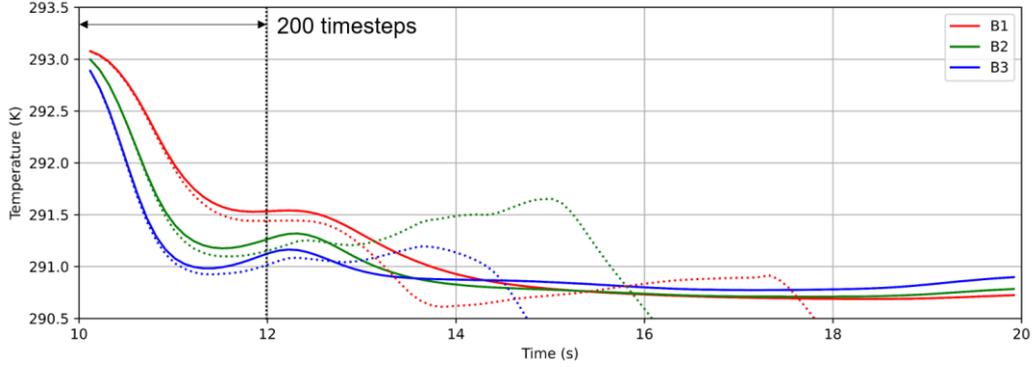

**Fig. 1.** Results of temperature time series prediction with single training approach. The sold line shows the CFD ground truth path and the dotted line shows the ML prediction path. Although the FVMN model shows good agreement in the initial extrapolation, it is vulnerable to long-term unsteady simulations.

## 2. Machine learning method

2.1. Residual-based physics-informed transfer learning (RePIT) strategy

In this study, we propose a novel ML strategy to accelerate CFD simulations – RePIT. Transfer learning was adopted in our strategy to address the error divergence problem in conventional single ML attempts. RePIT strategy can also be defined as the ML-CFD hybrid computation method. As shown in **Fig. 2(a)**, in the initial time series, this strategy starts CFD simulations identically to the general fluid simulations. Subsequently, the calculated CFD time series data are used to train the constructed neural networks. The number of time steps included in the first training dataset may depend on the complexity of the simulations. Now, the trained networks can rapidly predict subsequent CFD time series data without having to perform expensive first principles calculations. Neural networks, defined as universal nonlinear approximators, are much faster than conventional nonlinear solvers [24].

The main originality of this strategy is that one ML acceleration part ends before the error divergence problem occurs. In ML time series prediction, the error with the ground truth cannot be directly monitored since there is no ground truth simulation data. Therefore, the residuals of the ML-predicted time series were monitored (physics-informed manner), and one ML part ends when the residuals approach the tolerance level. Here, a residual is the magnitude of the conservation error in the governing equations, as defined in a typical CFD solver. The intermediate CFD simulations calculate the variable fields of the next time step while again reducing the conservation error below the tolerance level. These latest CFD results are used as training datasets to update the pretrained neural network parameters (transfer learning). The evolved neural networks again accelerate the CFD simulations by calculating the multistep data (another ML acceleration part); this physics-informed transfer learning process is repeated until the calculations are completed.

This residual-based transfer learning strategy seems reasonable because typical CFD solvers also calculate continuous time series based on an allowable residual criterion. The conventional CFD and this strategy are the same in that simulation stability is estimated based on the residual amounts; the only difference is whether the first principles are calculated with FVM or neural networks. In fact, recent engineering studies classify neural networks as one kind of numerical methods such as FVM [23]. To the best of the authors' knowledge, this is the first study to apply the residual-based ML-CFD hybrid computation method to accelerate heat and mass transfer simulations. The key advantage of this transfer learning strategy is that it can be individually applied to various simulation because the required initial training data is negligible. Although some network optimization guidelines are required depending on the simulation complexity, network training using numerous simulation results is unnecessary. In other words, it is not required to reproduce a huge dataset considering variations in geometry and boundary conditions.

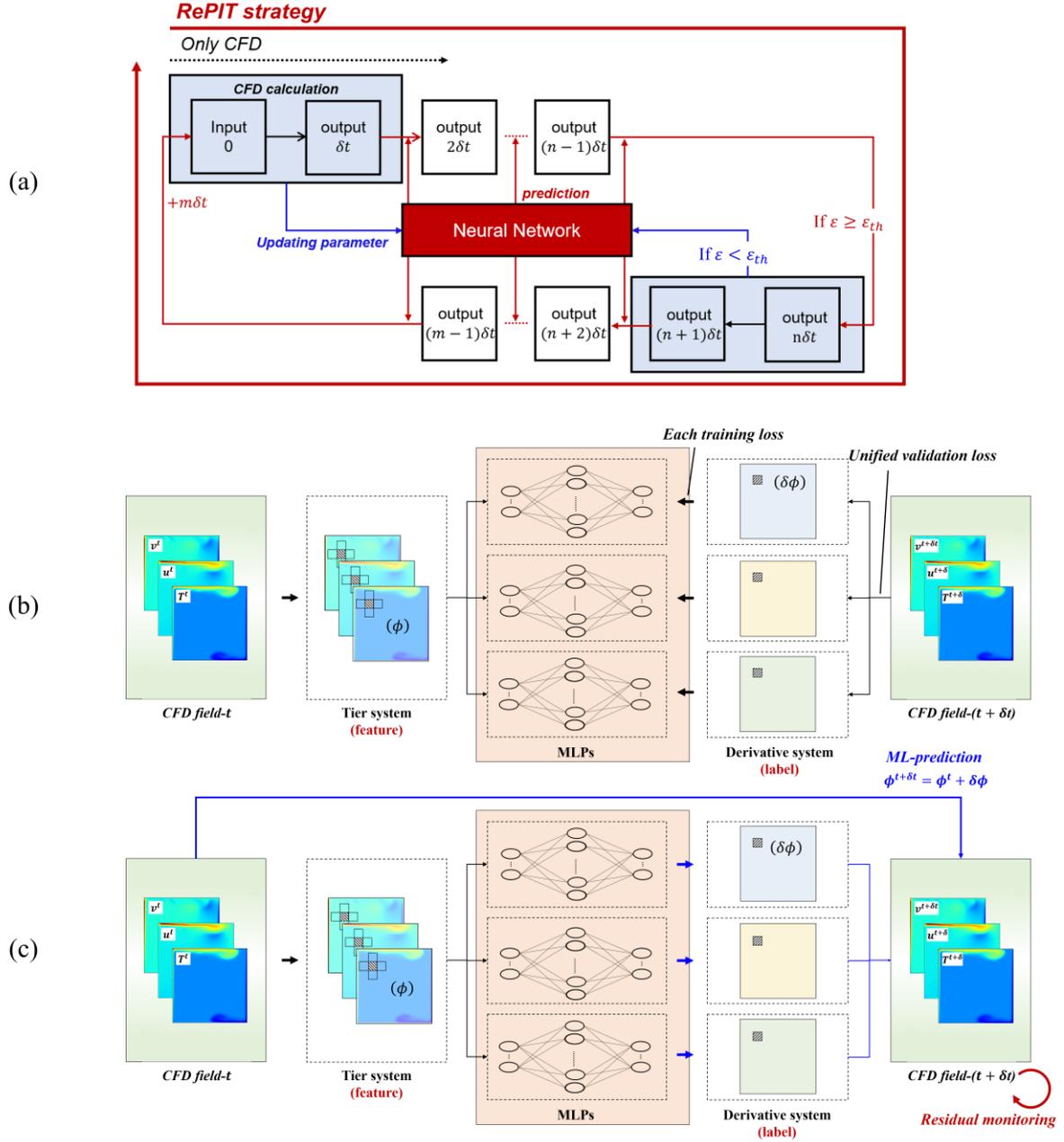

**Fig. 2.** (a) Framework of RePIT strategy to efficiently accelerate CFD simulations. Cross point can be pinpointed in unsupervised manner by calculating residuals of conservation equations. (b) Training and (c) prediction procedures of FVMN with tier-derivative system (same as network model in **Fig. 12(a)** of Ref. [17], but more detailed version). FVMN model, which introduces FVM principles, and residual calculation algorithm are combined to reinforce physics-informed strategy.

To implement and verify RePIT strategy, an effective neural network model is necessary to accurately predict continuous time series data with a small amount of CFD time series training data. The network type, architecture, and loss function significantly affect the network speed and accuracy for time series predictions. Convolutional neural networks (CNNs) are unsuitable for this hybrid computation method because they require numerous images as the training data [19, 20]. Concurrently, many recent studies have reported the strong performance of recurrent neural networks (RNNs) and long short-term memory (LSTM) in time series prediction [26, 27]. The objective of these recurrent models is to predict a future time series from an earlier time series data while minimizing information loss [26, 28]. However, the efficacy of these recurrent models is unsuitable for this strategy, which requires rapid updating of network parameters through short intermediate CFD simulations.

On the other hand, a grid-based neural network, which can be trained using a very small amount of time series data, such as a finite-volume method network (FVMN) model, fits well with this strategy [17, 29]. For this reason, the reliability of this strategy, and whether it is feasible to prevent error divergence through residual control, was verified using a grid-based neural network model in this study. Second, it is essential to develop a computational coupling method for ML and CFD. For the free input and output of CFD data without data loss, the OpenFOAM code, which is an open-source CFD software [30], was used in this study. The applied neural networks are described in the following sections and the coupling method are described in **Fig. S1** (in Supplementary materials).

2.2. Finite-volume method network

Most commercial CFD codes and the OpenFOAM code numerically solve the governing equations using a finite-volume method (FVM). Although previous studies have predicted CFD time series data using ML algorithms [29, 31], they have not sufficiently considered the characteristics of the FVM in their network models. A general grid-to-grid multilayer perceptron (MLP) can predict data in the training dataset; however, such a network has limitations when predicting data beyond the training set, and ultimately, future time series data. CNNs, which perform very well in computer vision, have been employed in flow-field predictions [32]; however, a considerably large number of images is required to determine the parameters of the kernels, as discussed in the previous section. For example, recent CNN studies used approximately 100,000 and 500,000 field images as training datasets for steady and unsteady flow field predictions, respectively [19, 20]. CNNs significantly improve image prediction accuracy by a reasonable assumption of the stationarity of statistics [33]. However, it should be noted that grids in a CFD snapshot are more strongly related to each other than pixels in a normal image due to the stationarity of the governing equations. In other words, a unique characteristic of a CFD field image is that all grids interact with neighboring grids using the same governing equations. Similar to the replacement of conventional MLP approaches in image processing with CNN ones with the stationarity of statistics assumption, the characteristics of CFD data allow the development of a more fitted network model.

Recently, Jeon et al. developed an FVM network (FVMN) model to predict a CFD time series by introducing FVM principles into the network architecture and the loss function [17]. FVMN model employs tier-input and derivative-output systems in the network architecture. In previous models, the CFD time series data at times $t$ and $(t + 1)$ are used as the input and output variables in the training process, respectively [29, 31]. The parameters (weight and bias) in the neural networks are iteratively updated using a backpropagation algorithm until the loss function converges [34]. Specifically, the parameters are simply determined by minimizing the mean square error (MSE) loss of the predicted values, $\phi(\theta)$, and the ground-truth values, $\phi$, as expressed in **Eq. (1)** [31]. The validation loss for the 20% dataset is monitored to prevent overfitting.

$$J(\theta) = \frac{1}{n}\sum_{k=1}^{n}\left\{\left(\left(\phi^{t+\delta t}\right)^k - \left(\phi^{t+\delta t}\right)^k(\theta)\right)^2\right\}. \tag{1}$$

In contrast, in the abovementioned FVMN model, a derivative-output system is designed for the output variables [17]. In the fluid flow analysis, the scales of the actual variable values in the previous time step are significantly larger than the change in the magnitude up to the next time step. Thus, accurately predicting the change using an actual value-based loss function is difficult. Note that, in other engineering fields, several neural network studies have successfully improved the performance by the scale separation of the output variables [35]. Jeon et al. confirmed the significant effect of scale separation on improving the network performance, using a reacting-flow CFD dataset and by simultaneously conducting reaction and radiation calculations [17]. Consequently, the parameters are learned by minimizing the MSE loss of the variation in each variable, $\delta\phi_i = (\phi_i^{t+\delta t} - \phi_i^t)/\delta t$, as expressed in **Eq. (2)** [17]. Note that, $i$ denotes the variable type, and the time derivative is included for brevity in the following equations. Moreover, a unified validation loss for checkpoints (**Eq. (3)**) is applied to implement a physics-informed loss function [17].

$$J_{train,i}(\theta) = \frac{1}{n_{train}} \sum_{k=1}^{n_{train}} \left\{ \left((\delta\phi_i)^k - (\delta\phi_i)^k(\theta)\right)^2 \right\}, \quad (2)$$

$$J_{val}(\theta) = \frac{1}{n_{val}} \sum_{i=1}^{I} \sum_{k=1}^{n_{val}} \left\{ \left((\delta\phi_i)^k - (\delta\phi_i)^k(\theta)\right)^2 \right\}. \quad (3)$$

Additionally, a tier-input system allows the network model to solve the mass and heat transport among neighboring cells, which is considered the most important process in an FVM. The physical quantities at point $\phi_{i,j}$ and its neighboring grid points $\phi_{i-1,j}, \phi_{i+1,j}, \phi_{i,j-1}$, and $\phi_{i,j+1}$ are included in the input matrix, whereas previous MLP models only include $\phi_{i,j}$. Jeon et al. confirmed that their FVMN approach, which embodies the numerical principles of the CFD solver, has a noticeable advantage of increasing the accuracy of time series data prediction. The total number of training/validation examples, $n$, is identical to the number of grid points in the computational domain (grid-based training approach). As discussed in *Section 1*, grids in a CFD snapshot are more strongly related to each other than pixels in a normal image owing to the stationarity of the governing equations. Therefore, different from an image-based training approach, which requires numerous images as training examples (e.g., CNNs), this grid-based training approach can train the network very efficiently with a small amount of time series data.

Interestingly, as expressed in **Eq. (4)**, Jeon et al. also developed a physics-informed loss function that calculates the balance of each governing equation by the summation of the physical fluxes to further assimilate FVMN model to the FVM [17]. The residual, $\varepsilon_j^k$, for each grid $k$ and governing equation $j$ is iteratively calculated to determine the parameters in a mini batch. The number of governing equations (e.g., continuity and Navier–Stokes equations) depends on the simulation conditions, where $w_1$ and $w_2$ are the weighting factors used to adjust the scale of each function. The unified validation loss is the same as in **Eq. (3).** In their case study, the total residual converged to a lower value without compromising the training/validation error when the residual term was included in the loss function. Interestingly, the magnitude of the total residual was even smaller than that of a network trained by a double dataset size. Thus, providing the governing equation information in the loss function is as effective in increasing the network performance as increasing the amount of the training data. In this study, the residual amount was not included in the loss function because the hybrid computation method is the main objective. **Figs. 1(b) and (c)** show the FVMN framework during the network training and prediction stages of this study. A more detailed description of FVMN, including the physics-informed loss function, can be found in Ref. [17].

$$J_{train,i}(\theta) = \frac{w_1}{n_{train}} \sum_{k=1}^{n_{train}} \left\{ \left((\delta\phi_i)^k - (\delta\phi_i)^k(\theta)\right)^2 \right\} + \frac{w_2}{n_{center}} \sum_{j=1}^{J} \sum_{k=1}^{n_{center}} \varepsilon_j^k. \tag{4}$$

2.3. Residual monitoring

The residual computed by the CFD solver is the imbalance in the governing equations summed over all the computational cells. Regarding the residual as an error indicator has proved to be reasonable and efficient [36]. It was confirmed that the reduction in the residual by using the physics-informed loss function contributes to the improvement in the prediction accuracy. However, a more notable finding was that the residuals calculated in an unsupervised manner can monitor the network accuracy in an ML acceleration part [17]. Therefore, we postulated that the error trend can be monitored by calculating the residuals of the conservation equations by the same method as that of first-principles calculations. This optimal use of the physics-informed loss function as a monitoring function makes our RePIT strategy feasible.

**Eqs. (5) and (6)** present the conservation equation for a general variable $\phi$ at a grid point $P$ after discretization, where $a_p$ is the center coefficient, $a_{nb}$ is the neighboring grid-point coefficient, and $b$ is the contribution of the source term [37].

$$a_P \phi_P = \sum_{nb} a_{nb} \phi_{nb} + b, \tag{5}$$

$$a_p = \sum_{nb} a_{nb} - S_p. \tag{6}$$

As expressed in **Eq. (7)**, in general CFD solvers, the average residual can be calculated as the sum of the imbalance amounts over all computational grid points [37]. However, determining convergence by observing the unscaled residuals calculated using **Eq. (7)** is difficult because there is no explicit standard. Therefore, the solvers employ the initial residual values as the standard [37]. In this study, the residual values in the ML-predicted time series were also monitored using scaling based on the residual values in the training dataset. It is worth noting that OpenFOAM solvers also check convergence based on the scaled residual using the average of the solution vector [30]. The detailed residual equations designed according to the CFD modeling in this study are introduced subsequently in *Section 4*.

$$R^\phi = \sum_{cells\ P} |\sum_{nb} a_{nb} \phi_{nb} + b - (\sum_{nb} a_{nb} - S_p)\phi_P|. \tag{7}$$

## 3. CFD dataset

3.1. Modeling and simulation

To investigate the feasibility of RePIT strategy, we generated a two-dimensional natural convection flow CFD dataset for a case study. It should be noted that most prior studies have verified their ML models with the flow field simulation excluding heat transfer equation [17, 19, 20] , but most of CFD applications include heat transfer. The natural convection flow was induced by a uniform wall temperature of hot and cold walls. For this simulation, the OpenFOAM solver buoyantPimpleFoam, which is an unsteady solver for buoyant-driven turbulent flow of compressible fluids for heat transfer was employed [38]. This solver was validated by Neilsen's airflow experimental results with heated wall [39, 40]. To solve the velocity, pressure and energy equations, the buoyantPimpleFoam solver uses the (PIMPLE) algorithm, which is a combination of PISO (Pressure Implicit with Splitting of Operator) and SIMPLE (Semi-Implicit Method for Pressure-Linked Equations). The continuity, momentum equation and energy equation solved using buoyantPimpleFoam are expressed in **Eqs. (8-10)**, where $\rho$ is the density, **u** is the velocity vector, $p$ is the static pressure, $\mu_{eff}$ is the effective viscosity (sum of molecular and turbulent viscosity), $h$ is the sum of the internal energy, $K$ is kinetic energy, and $\alpha_{eff}$ is effective thermal diffusivity (sum of laminar and turbulent thermal diffusivities). In this study, the air with a perfect gas assumption modelled as the working fluid. Because there was a negligible pressure change across the entire domain in this simulation, the fluid density was actually determined by temperature, similar to the incompressible solver. The source code of OpenFOAM buoyantPimpleFoam including the governing equation is available at https://github.com/OpenFOAM/OpenFOAM-2.1.x/blob/master/applications/solvers/heatTransfer/buoyantPimpleFoam/buoyantPimpleFoam.C and a more detailed description can be found in Ref. [38].

$$\frac{\partial \rho}{\partial t} + \nabla \cdot (\rho \mathbf{u}) = 0, \tag{8}$$

$$\frac{d\rho \mathbf{u}}{dt} + \nabla \cdot (\rho \mathbf{u}\mathbf{u}) = -\nabla p + \rho g + \nabla \cdot \left(\mu_{eff}(\nabla \mathbf{u} + \nabla \mathbf{u}^T)\right) - \nabla \left(\frac{2}{3}\mu_{eff}(\nabla \cdot \mathbf{u})\right), \tag{9}$$

$$\frac{d\rho h}{dt} + \nabla \cdot (\rho \mathbf{u} h) + \frac{d\rho K}{dt} + \nabla \cdot (\rho \mathbf{u} K) - \frac{\partial p}{\partial t} = \nabla \cdot (\alpha_{eff} \nabla h) + \rho \mathbf{u} \cdot g. \tag{10}$$

To generate the geometry and mesh for the natural convection simulation, we used the blockMesh utility, which supplied with OpenFOAM. As shown in **Fig. 3**, a simple rectangular 2D geometry was employed as the computational domain for this study. The rectangular geometry represents closed channel, consisting of one hot wall, one cold wall, and two adiabatic walls. The hot wall positioned on the left side of the channel, and cold wall was located opposite side of the hot wall. The adiabatic walls were defined as the top and bottom wall. The mesh was constructed using uniform 40,000 hexahedral cells. An overall mesh quality inspection confirmed that the maximum nonorthogonality was zero, and that the maximum skewness was close to zero. To induce the natural convection in the channel, the following calculation conditions were set. The temperature of how wall set as 307.75 K, the cold wall set as 288.15 K, and the initial internal air set as 293 K. The internal pressure was 1 atm, and velocity fields were initialized to zero. The total simulation time and the time step were 100 s and 0.01 s, respectively.

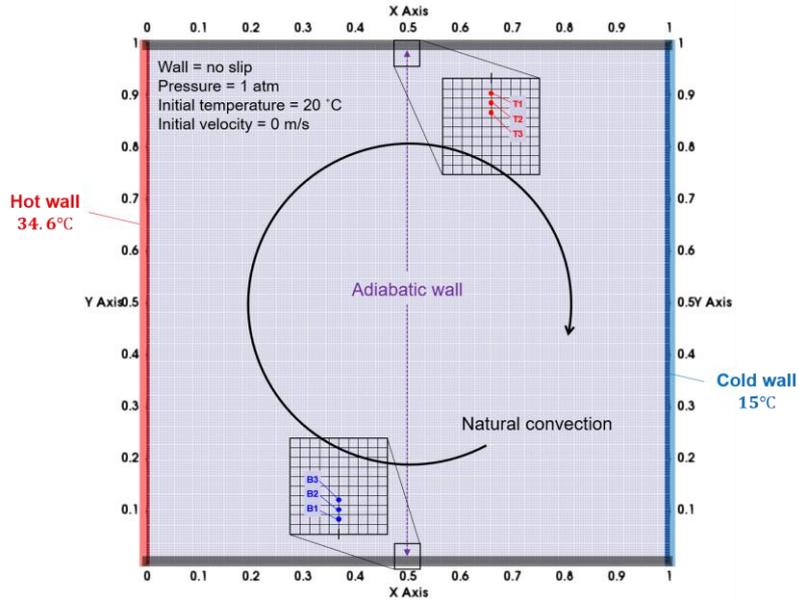

**Fig. 3.** A schematic showing the geometry, grids and boundary conditions of natural convection simulation. Three points near the top wall (T1-3) and three points near the bottom wall (B1-3) were used for local variable accuracy evaluation in *Section 4*.

3.2. Dataset

**Multimedia data. S1** (in Supplementary materials) shows the full video obtained from the aforementioned simulation. At the hot wall, the air whose temperature has increased rises due to buoyancy force. Conversely, the air whose temperature has decreased on the cold wall descends. In other words, a natural convection flow is formed due to the difference in density. The rose and descended flow from the hot wall and cold wall transfer to the opposite wall gradually during 0 to 10 s. At the hot wall, because the temperature gradient with initial air temperature was larger than cold wall, the convection flow originating from hot wall is faster than cold wall. Consequently, the convection flow from the hot wall collides with cold wall near the 10 s and induce the mixing near the cold wall. After about 8 s from the first mixing near the cold wall, the cold flow originating from the cold wall collides with the hot wall. Subsequently, the continuous clockwise flow is formed inside the channel. Because natural convection has a long flow timescale compared to forced convection, the flow inside the channel begins to stabilize over time with oscillations by mixing.

To investigate the local variations, three points near the top wall (T1-3) and three points near the bottom wall (B1-3) were observed as shown in **Fig. 3**. **Fig. 4** shows the temperature and velocity time series data at each location. During the initial period of natural convection flow formation (0 -10 s), the x-axis velocity and temperature of T1-3 sharply increased when the heated flow reached the probe locations. It is important to note that, due to the boundary viscous force near the top wall, point T3 exhibits the highest velocity, while point T1 exhibits the lowest velocity. However, the temperature of T1-3 shows a similar value. At the points near the bottom wall, the x-axis velocity of B1-3 increased, and the temperature sharply decreased when the flow from the cold wall reached the points. After the first collision of the heated flow with the cold wall (10 s), the velocity value began fluctuating, and the reverse flow generated due to mixing. After 20 s, the fluctuations gradually subsided over time. Because the 10 -20 s interval exhibited the most irregular and rapid changes, this interval's time series was used to verify RePIT strategy.

The input and output variables for FVMN model were extracted from the ML domain (the number of training examples is n = 198 × 198 per a time series data), similar to the method in Jeon et al.'s study [17]. We assumed that the grids of the first boundary layer excluded from the ML domain are computed from the CFD solver. Because ML predicts all values except for the first boundary layer cells, the variables of those cells can be calculated with only a few iterations with the boundary condition. An acceleration performance comparison of the method of including all grids in the ML zone by increasing the network model/training data and the method of including the CFD solver in the boundary region is our future work.

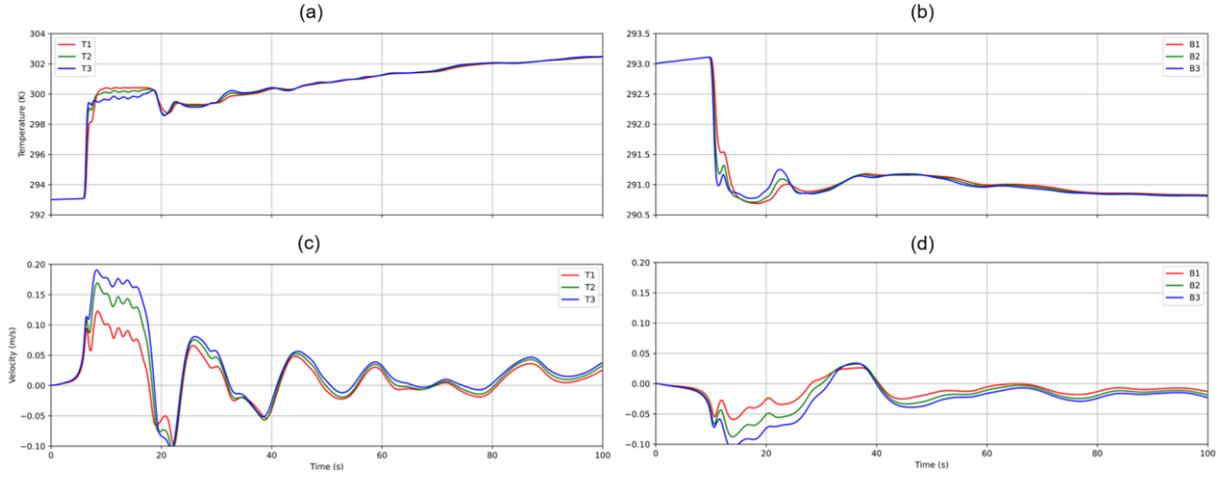

**Fig. 4.** CFD Time series data at specific locations (each location is marked in **Fig. 3**). The natural convection begins to stabilize after 40 s. In this study, RePIT strategy was validated with 10 -20 s temperature time series data with the largest fluctuation in *Section 4*.

## 4. Feasibility study of RePIT strategy

4.1. Implementation of single training approach and RePIT strategy

In this study, both the single training approach and the RePIT strategy used FVMN model. Neural networks, such as FVMN, contain hyperparameters that can significantly affect the network's accuracy and speed performance (e.g., network architecture, loss function, number of layers, number of neurons, and activation function). Because these hyperparameters cannot be determined during the training process, it is necessary to optimize them in advance. The FVMN hyperparameters were optimized based on the absolute error in the training/validation dataset following the method described in Ref. [17]. Consequently, 11 hidden layers, 512 nodes per layer, and the rectified linear unit function (ReLU) were employed. Many studies have reported that the degradation of network performance despite increasing the network size is a typical problem in neural network methods, which is called the overfitting problem [41]. In this study, overfitting was identified in the validation dataset when the number of hidden layers and nodes was increased. The Adam optimizer was used to iteratively update the parameters during the training stage [42]. To confirm the reliability of the constructed network model, the variation in the validation loss function during the training procedure was examined. It was observed that the loss function gradually reduces with the epoch number and converges well after 5000 epochs.

As discussed in *Section 3*, this study evaluated the feasibility of RePIT strategy based on a 10 – 20 s region of the natural convection simulation, which represents the most chaotic region. The prediction accuracy was investigated both as changes in local variables (B1-3, T1-3 in *Section 3*) over the entire time period and in domain-wide variable distributions at each timestep. FVMN model employs a grid-based training method, allowing it to predict future time series using only the first few datasets. In the single training approach (*Section 4.1*), three time series data (10, 10.01, and 10.02 s) were used for training. The training set consisted of 80% of the data from 10, 10.01, and 10.02 s, while the remaining 20% was used for validation (input: $\phi^{10}, \phi^{10.01}$, output: $\delta\phi^{10.01-10}, \delta\phi^{10.02-10.01}$). This grid-based training approach resulted in a total of 79,202 training samples for each variable, excluding the grids of the first boundary layer, as previously mentioned. For a fair performance comparison, the RePIT strategy's initial training used the same three time series data (10, 10.01, and 10.2 s) and the same training method as the single training approach. However, unlike single training approaches, RePIT strategy periodically updates network parameters using new CFD time series in the hybrid computational framework. Considering the usual CFD iteration method which sets the allowable residual size on a logarithmic scale, the maximum allowable residual size was set to five times the value obtained from the OpenFOAM ground truth data. Optimization of the maximum residual size for acceleration performance is a topic for future work. The ML-CFD hybrid computation process is described in more detail in *Section 4.3*.

4.2. Limitation of single training approach

In this section, we evaluate the accuracy of the single training approach by comparing with the ground-truth data (original CFD data). **Fig. 5** illustrates the variations in temperature and x-axis velocity when predicting continuous time series data with the network trained using the 10, 10.01, and 10.02 s datasets. Remarkably, for all locations, the FVMN model, trained with only the initial three time series data, succeeds in extrapolating up to 200 timesteps (12 s) and even predicts the temperature and velocity behavior quite well after the inflection point. This outcome is noteworthy as the neural network was never trained with time series data beyond the inflection point, demonstrating that the FVMN model acts as a nonlinear function approximator for CFD governing equations in unseen data. However, the error gradually increases with time, and after 14 s, the velocity at the B3 location exhibits a tendency that is completely opposite to the ground truth.

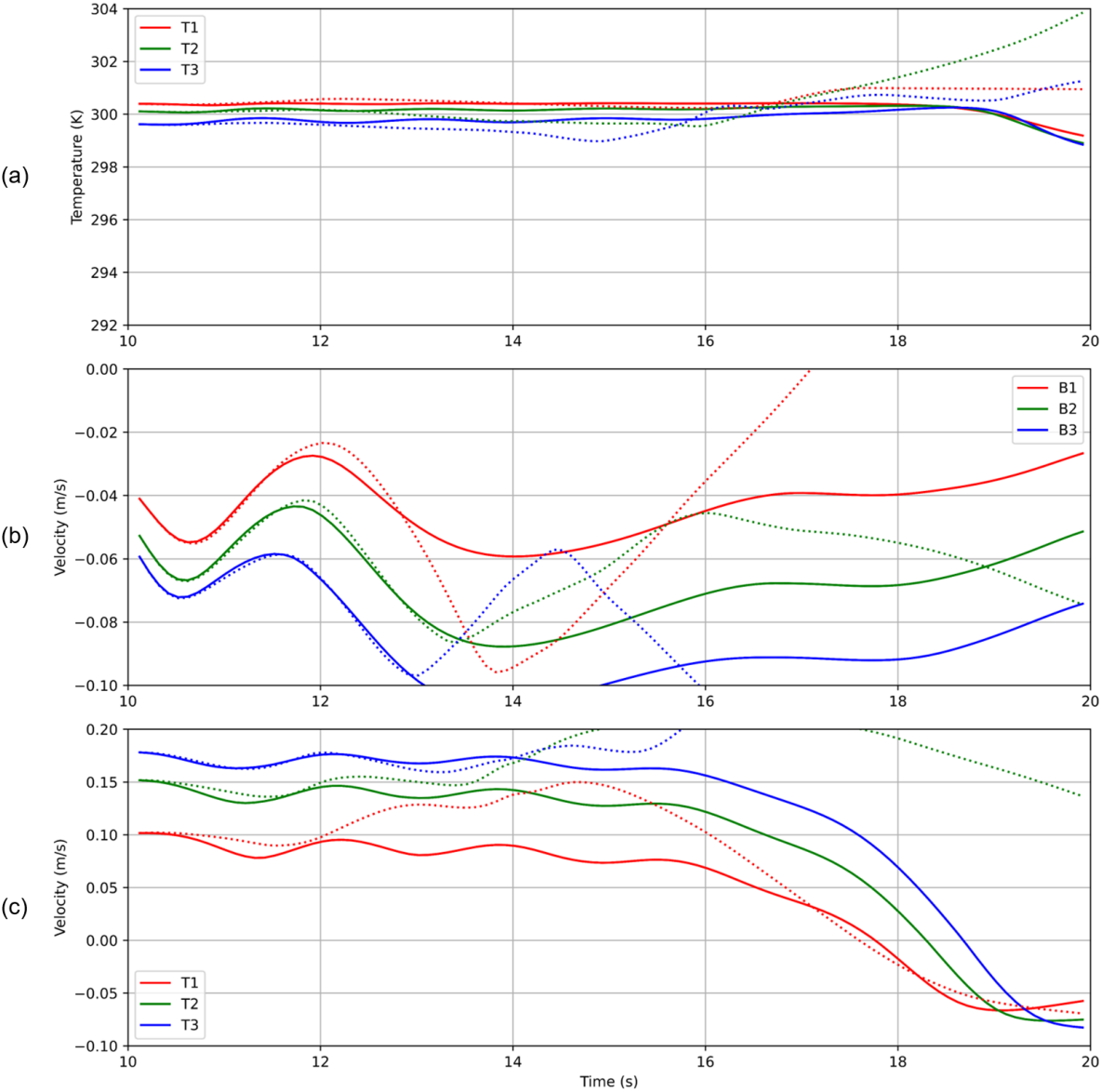

**Fig. 5.** Results of time series prediction with single training approach. (a) temperature at location T1-3. (b) x-axis velocity at location B1-3. (c) x-axis velocity at location T1-3 (temperature at location B1-3 was shown in **Fig. 1**). The sold line shows the CFD ground truth path and the dotted line shows the ML prediction path. All values begin to diverge after 12 s.

Previous studies have revealed that completely suppressing the increase in error as the interval between the training and prediction times increases is unrealistic in the single training approach [19]. In CFD simulations, the residual of each conservation equation is prevented from exceeding the allowable range using an internal iteration algorithm at each time step [37]. However, in the ML-predicted time series, the residuals increase because the neural network cannot converge the residuals of the governing equations with the pressure field projected by the inner iterations. As shown in **Fig. 6**, the scaled residuals in the ground-truth time series remain almost constant at 1.0 for the continuity, momentum, and heat equations, respectively. At this time, the residuals were scaled based on the residuals from the 10 s ground-truth data.

On the other hand, in the ML-predicted time series, the residual of the continuity gradually increases, and a scale change occurs before 100 timesteps. Similarly, the residual of the heat transfer equation also undergoes a scale change near 200 timesteps. In CFD simulations, a change in the residual scale serves as an indicator of a convergence problem occurrence. When the calculated variable distribution no longer follows the governing equations, the error from the ground truth rises rapidly. This is the reason why the temperature and velocity contour at 20 s shows non-physical pattern as well as a large difference from the ground truth (**Fig. 7**). In summary, treating a long-term time series using this single training approach is difficult because the growing residual problem eventually cause the increasing error with the ground truth.

To describe the error (residual) divergence process in hydrodynamic terms, we should pay attention to the near-wall flow, especially the top wall. In viscous flow, the velocity of a grid is greatly affected by the velocity of surrounding grids because viscous force is generated due to the velocity difference with surrounding grids. As shown in **Fig. 5(c)**, the first variable for which the FVMN model predicted value deviated from the CFD trajectory was the velocity at T1 (around 10.5 s). This time is almost the same as the time when the continuity residual undergoes a scale change. Because T1 is closest to the top wall, it is most affected by the bias between the first grid layer velocity calculated by CFD and the velocity from the second layer calculated by FVMN. In other words, the residual was primarily generated from the boundary layer flow calculation. This problem occurred because, unlike CFD solvers, it was difficult to impose boundary conditions in data-driven methods. This is another reason why our hybrid method with residual recovery is required. Compared to the top wall, the velocity range and scale near the bottom wall were smaller and therefore less affected by biased data. The error generated in the T1 velocity field was propagated to the nearby velocity field (T2, T3) through a network solving the continuity and Navier-Stokes equation and eventually to temperature errors through a network solving the heat equation.

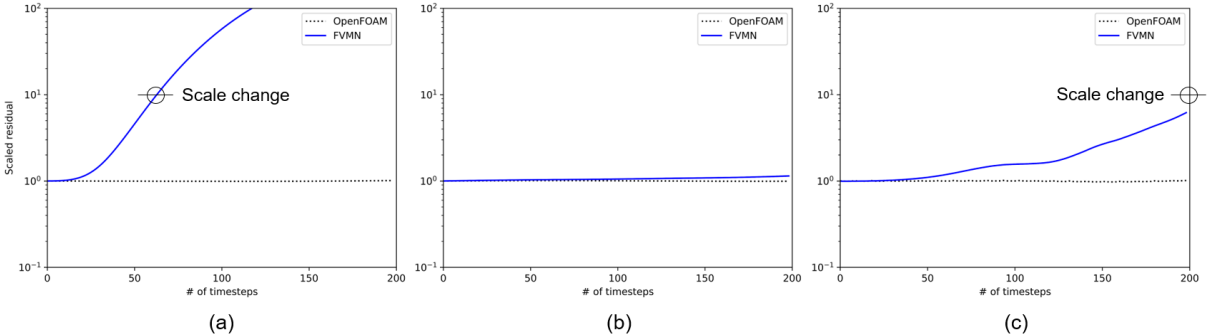

**Fig. 6.** Residuals of predicted time series with single training approach. (a) continuity equation (b) Navier-Stokes equation (c) heat equation. It was confirmed that variable fields violate the governing equation over time.

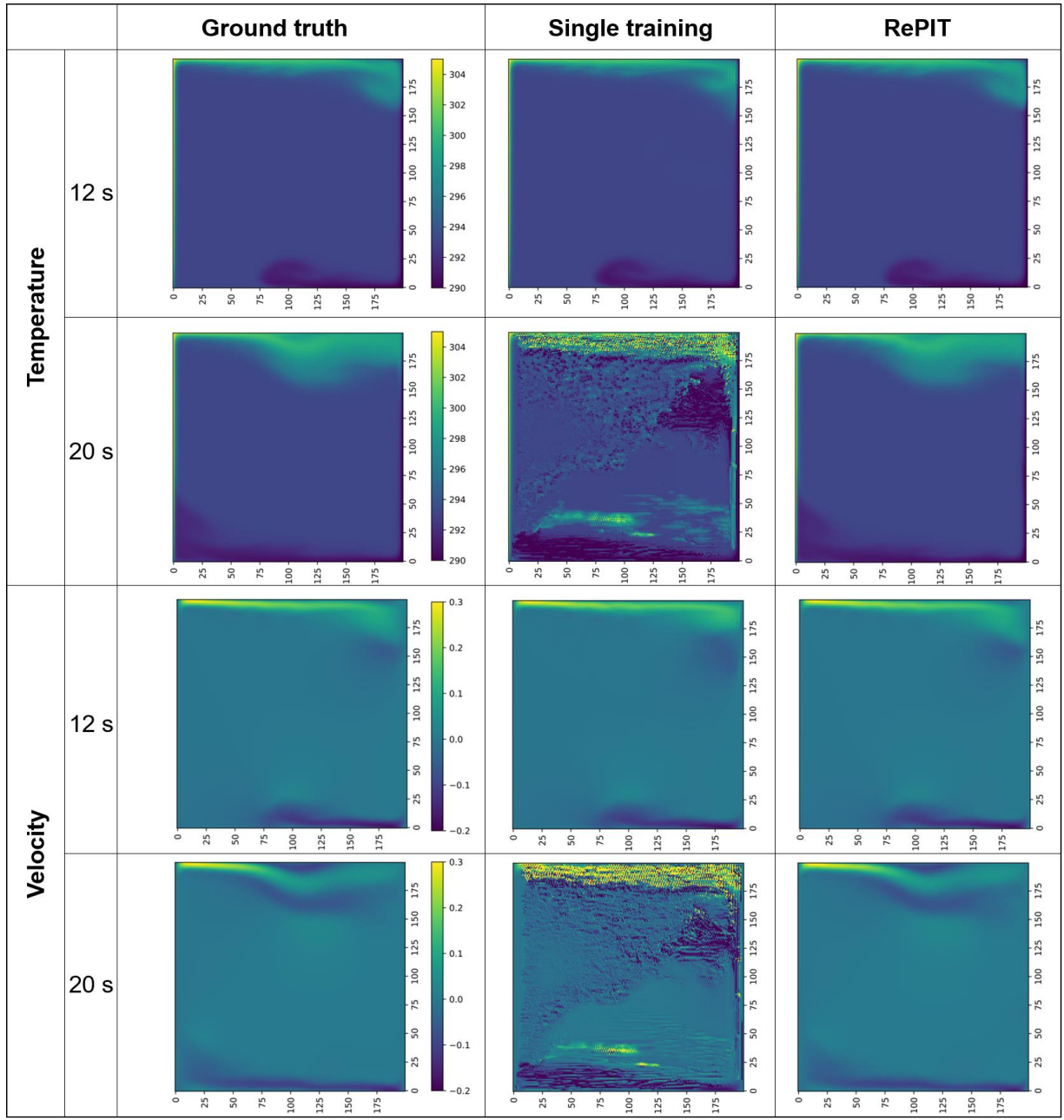

**Fig. 7.** Comparison of velocity contours between the ground truth, single training approach and RePIT strategy. Contours of the single training at 20 s which does not satisfy the first principles shows non-physical pattern. On the other hand, contours of RePIT strategy continues to agree with the ground truth.

4.3. Performance of RePIT strategy

The key concept of RePIT strategy is to consistently prevent residuals in the ML-predicted time series from exceeding the tolerances by conducting intermediate CFD simulations. Our hypothesis is that long-term CFD simulation is possible with our hybrid method if not only the residuals of the governing equations can be kept below the tolerance levels but also if the network parameters can be updated with the analogous variable topology in the latest CFD time series data (transfer learning manner). As shown in **Fig. 2(a)**, the intermediate CFD simulations aim to predict the next time series while matching the magnitudes of the left and right terms of each conservation equation. This residual-based strategy seems reasonable because general CFD simulations also determine whether the next time step can be calculated based on the sizes of the residuals. For example, the residuals of the continuity, momentum, and heat transfer equations of the CFD dataset in this study were also maintained at a set scale. Regarding the residual as an error indicator have proved to be reasonable and efficient [36]. As shown in **Eq. (7)**, the residual can be computed by imbalance in the governing equation form. Unlike mathematical principles, the residual cannot be 0 in actual computation, so **Eq. (8-11)** is change to **Eq. (11-13)** during the internal iteration [30]. The internal iteration in the CFD algorithm can reduce the residual of the momentum equation, including the momentum change in the grid and the inflow/outflow momentum, to an acceptable range ($\mathrm{argmin}_{p^*,\mathbf{u}^*,T^*} \sum_n \varepsilon$) [30]. The continuity and heat transfer equations are solved with the same process (**Eq. (12, 13)**) for calculating the next time series ($u^*, v^*, T^*$). In this section, we discuss the efficacy of the intermediate CFD simulations for residual and error reductions.

$$\frac{\rho^* - \rho_{ML}^t}{\delta t} + \nabla \cdot (\rho \mathbf{u})^* = \varepsilon, \tag{11}$$

$$\frac{(\rho \mathbf{u})^* - (\rho \mathbf{u})_{ML}^t}{\delta t} + \nabla \cdot (\rho \mathbf{u}\mathbf{u})^* + \nabla p^* - \rho^* g - \nabla \cdot \left(\mu_{eff}(\nabla \mathbf{u} + \nabla \mathbf{u}^T)\right)^* + \nabla \left(\frac{2}{3}\mu_{eff}(\nabla \cdot \mathbf{u})\right)^* = \varepsilon, \tag{12}$$

$$\frac{(\rho h)^* - (\rho h)_{ML}^t}{\delta t} + \nabla \cdot (\rho \mathbf{u} h)^* + \frac{(\rho K)^* - (\rho K)_{ML}^t}{\delta t} + \nabla \cdot (\rho \mathbf{u} K)^* - \frac{p^* - p_{ML}^t}{\delta t} - \nabla \cdot \left(\alpha_{eff} \nabla h\right)^* - \rho^* \boldsymbol{u}^* \cdot g = \varepsilon \ . \tag{13}$$

In this study, the ML-CFD cross is operated depending on the monitored continuity equation, as the scale change of the residual first occurs in the continuity equation. As shown in **Fig. 8**, the residual of the continuity equations gradually increases in the ML zone. The residuals were scaled based on the residuals from the 10 s ground-truth data. Once the residual reaches the set maximum residual criterion, it crosses into the CFD zone. In the CFD zone, the residual decreases and subsequently returns to the initial scale. On the other hand, the computation reenters the ML zone when the residual come back to the magnitude of the OpenFOAM original and simultaneously transfer learning is performed with the latest three time series data. It was noted that each transfer learning only required a very small number of 10 epochs compared to the initial network training (5000 epochs). This very short parameter updating time enables simulation acceleration through RePIT strategy. For example, as expressed in **Eq. (14)**, the parameters determined by training using the 10 –10.02 s dataset are first updated using the 10.61–10.63 s dataset $(\delta T^*, \delta u^*, \delta v^*)$ [34].

$$\theta_j := \theta_j - \alpha \frac{\partial}{\partial \theta_j}\left(\frac{1}{n_{train}}\sum_{k=1}^{n_{train}}\left\{\left((\delta\phi^*)^k - (\delta\phi^*)^k(\theta)\right)^2\right\}\right) \quad (14)$$

Most importantly, it was confirmed that RePIT strategy successfully recovered the residual every time for 1,000 timesteps (**Fig. 8**). On average, 20 ML prediction time series and 10 CFD simulation time series were repeated. This ratio may vary depending on the simulation complexity and the set maximum allowable residual size. In conclusion, 2/3 of the total simulation time period benefitted from neural net acceleration. Detailed calculation logs are described in **Table S1** (in Supplementary materials). The residual stabilization method was confirmed to work well, but the prediction accuracy must be verified together to confirm the efficacy of our acceleration strategy.

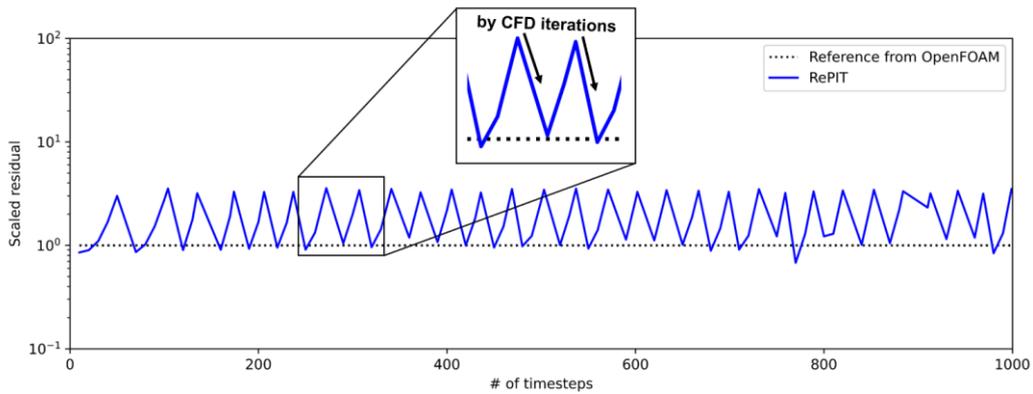

**Fig. 8**. Scaled residual of predicted time series with RePIT strategy. It was noted that the residual is well maintained within the set range, unlike the single training approach.

**Fig. 9** shows the efficacy of RePIT strategy based on the same time series data in *Section 4.2*. Before entering the second CFD zone, the calculation procedures for the single training and the RePIT strategy were the same. After that, the cross points ML→CFD and CFD→ML were pinpointed depending on the monitored residual and the set allowable range. Notably, the prediction path agreed well with the OpenFOAM ground truth path over the entire time period. The enlarged inflection point section of **Fig. 9(a)** also exhibits good accuracy. This good accuracy was confirmed in the whole domain, not in a specific region until the last timestep as shown **Fig. 7**. In the single training approach, the simulation error was primarily generated by biased velocity input between boundary layers. The error started to be noticeable in the T1 velocity field starting at about 10.5 s. On the contrary, in RePIT strategy, no errors that could affect the future time series calculation were identified at T1 as shown in **Fig. 9(d)**. In the CFD zone of the RePIT strategy, the residuals due to biased velocity input are continuously removed. We concluded that only the numerical constraint on the satisfaction of governing equations can solve the error accumulation problem in unsteady CFD simulations.

The difference between the predicted and the ground truth temperature is the largest at 16 – 18 s for the B1-3 and 12 -14 s for the T1-3. What these time series temperature trends have in common is that the temperature change rate with time is negligibly small. It seems that the effect of truncation error in OpenFOAM and TensorFlow data exchange procedures becomes noticeable when the change rate is quite close to zero. Nevertheless, we conclude that the feasibility of the long-term CFD simulation by RePIT strategy was validated in the natural convection case study. The maximum error from the ground truth near the top and bottom walls was below 0.4 K for temperature and 0.024 m/s for x-axis velocity. It is confirmed that the transfer learning strategy does not compromise the simulation accuracy for computational acceleration. Validation in the CFD case with more nonlinear and irregular governing equations such as turbulence, and reducing truncation errors through full computation framework integration, is our future work. All the Python/TensorFlow codes for RePIT strategy are available at https://github.com/41monster/RePIT_NaturalCirculation2D. The evaluation of the practical acceleration performance of the strategy is discussed in the next section.

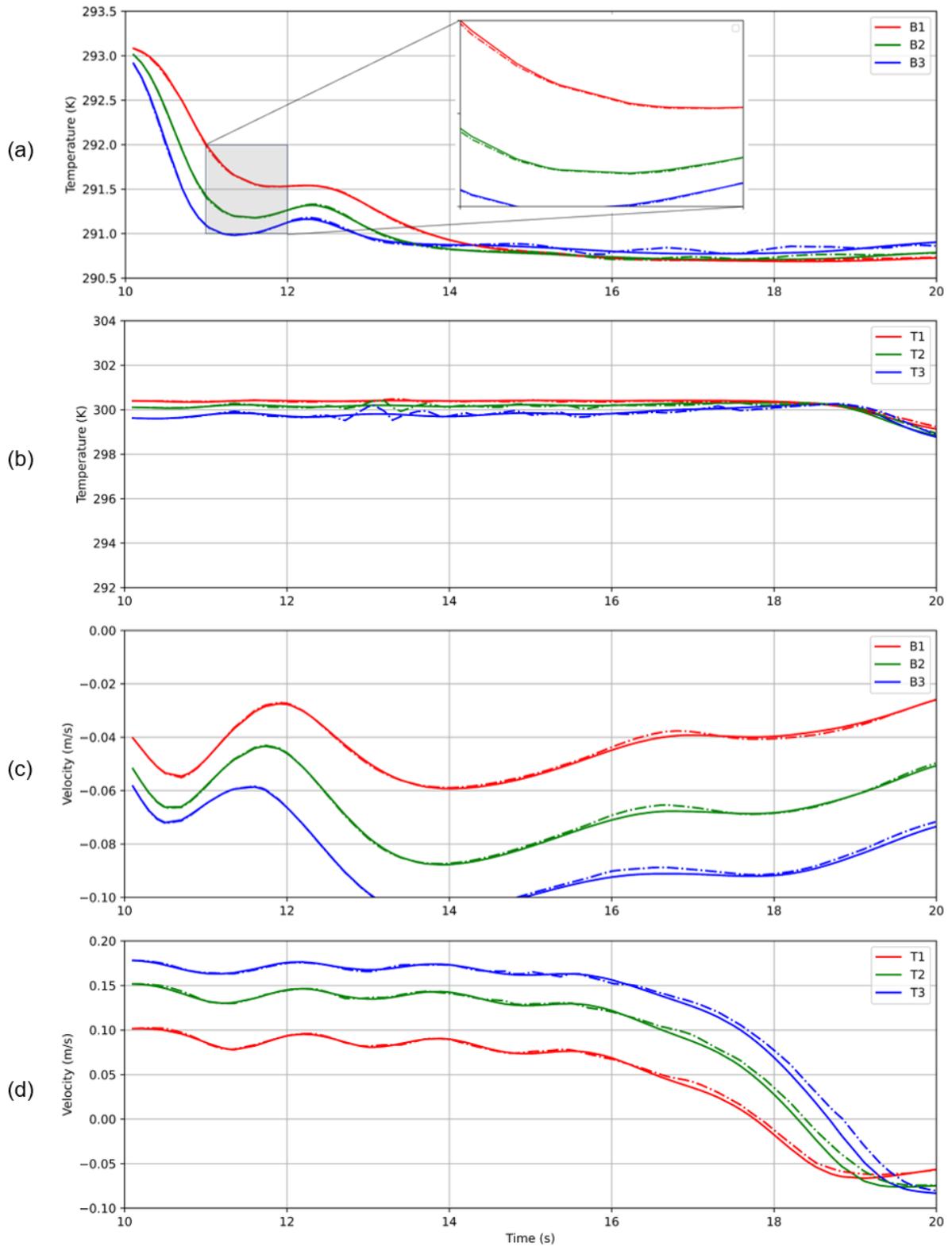

**Fig. 9**. Results of time series prediction with RePIT strategy. (a) temperature at location B1-3 (b) temperature at location T1-3. (c) x-axis velocity at location B1-3. (d) x-axis velocity at location T1-3. The sold line shows the CFD ground truth path and the dash-dotted line shows the RePIT prediction path. Significantly, the error accumulation problem was not observed unlike single training approach.

## 4.4. Speed comparison with general CFD solver

Although the long-term computational potential of RePIT strategy was verified, the acceleration performance should be evaluated to determine its practicality. **Fig. 10** compares the computational timelines of this strategy and general CFD simulations. The CFD simulation speed can be quantified as the average computation time for one timestep. However, in this strategy, four types of times comprise the total computation time: network training time ($t_{tr}$), parameter updating time ($t_{up}$), ML computation time for one timestep ($t_{ML}$), and CFD computation time for one timestep ($t_{CFD}$). Because RePIT strategy requires only a negligible amount of training data/time (3 time series in this study), in long-term unsteady simulations, the sum of the other three times ($t_{up}, t_{ML}, t_{CFD}$) represents the quantified simulation speed. The negligible amount of training data is the reason why RePIT strategy is a universal method which can respond to geometry or boundary condition changes. The parameter updating time (transfer learning time) is always required when transitioning from CFD simulations to ML calculations, and it is an important factor for evaluating the acceleration performance. In this study, we observed that the validation loss decreased to the same scale of the initial training stage with only 10 epochs in each transfer learning stage. In summary, the acceleration performance, $\psi$, in the hybrid computation can be evaluated using **Eq. (15)**. Note that $n_{CFD}$ and $n_{ML}$ are the numbers of ML and CFD time steps calculated in a bundle, respectively.

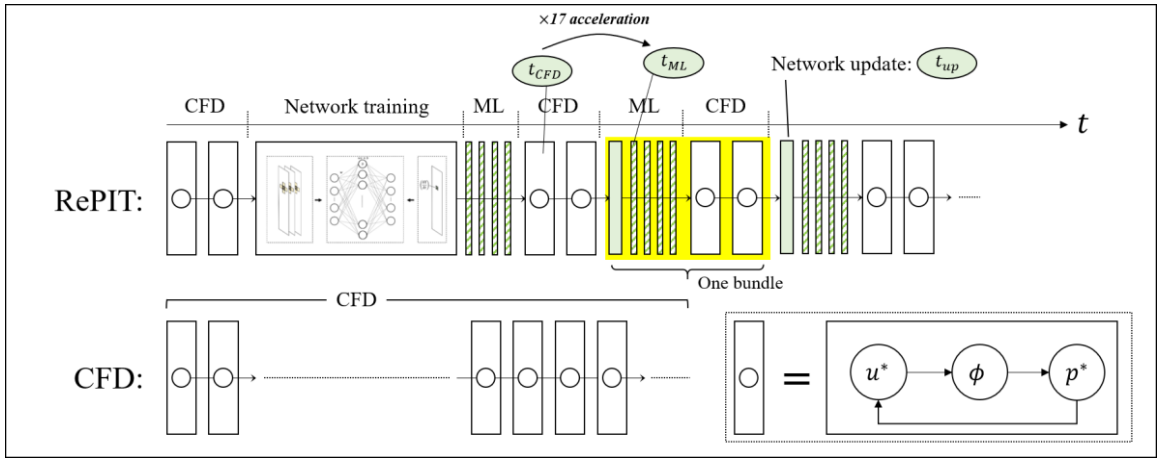

**Fig. 10.** Comparison of computational timelines of general CFD and RePIT strategy. In this strategy,

there are three time elements $t_{up}, t_{ML}$ and $t_{CFD}$ that make up one bundle.

$$\psi = \frac{n_{CFD} t_{CFD}}{n_{CFD} t_{CFD} + t_{up} + n_{ML} t_{ML}} . \tag{15}$$

We compared the average computational speeds for calculating 1 timestep using ML-TensorFlow and CFD-OpenFOAM. The average speed was calculated based on the simulation execution time excluding data transport. The CPU and GPU systems were an Intel Xenon Gold (56 cores/112 threads) 6258R processor (2.70 GHz) and NVIDIA TESLA A100 (CUDA core, 40 GB MEM), respectively. Two systems with similar monetary values were used to compare the computational speed (~USD 10,000). The average time for 1 timestep by the ML and CFD calculations were 0.171 s and 0.015 s, respectively. At this time, the OpenFOAM average time was determined through optimization of the number of threads (8 threads) and the ML average time was measured using the time module in Python. FVMN could compute the CFD time series 11 times faster than CFD solver. Thus, the simulation was accelerated by about 1.9 times in the assumed 2:1 ML-CFD cross computation considering the parameter-updating time (**Eq. (15)**).

The traditional machine learning approach (single training approach) and the hybrid approach (RePIT strategy) each have distinct advantages and disadvantages. Because the single training approach undergoes training only once, it can accelerate CFD simulations very quickly after the initial training. As described in *Introduction*, this is why previous single training approaches showed better acceleration performance than this hybrid method. However, the error accumulation problem in unseen data has not yet been solved in a single training approach. In other words, the error will eventually exceed the tolerance in long-term CFD simulations. This error accumulation problem is more evident in the condition of multiple governing equations. On the other hand, RePIT strategy has the advantage of being a realistic solution that can be applied to multiple governing equation and long-term CFD simulations.

Although RePIT strategy may not achieve noticeable acceleration performance in this CFD case study, its performance increases for larger domain simulations. Our previous study observed that when the number of grids increased by 25 times, the CFD average time for 1 timestep increased more than 10 times while the ML average time increased only 3 times [43]. Note that the simulation can be accelerated by about 3 times with the same trained model. The increasing complexity of the non-linear matrix computation in the CFD simulation also can increase the speed difference between CFD and RePIT strategy. One additional drawback of RePIT strategy is that feasibility in non-uniformed mesh conditions has not been examined. Consequently, the validated acceleration capability can contribute to future studies investigating performance variations following simulation type as well as network, residual tolerance optimization and mesh conditions.

## 5. Conclusion

Previous studies have revealed that the gradual error accumulation in unsteady CFD simulation is an inevitable limitation of the conventional single training ML approach. They highlighted that the development of ML strategies for long term CFD acceleration was a remaining issue in the fluid dynamics industry. In this study, RePIT strategy using the hybrid computation framework of ML and CFD was established. Our hypothesis is that long-term CFD simulation is possible with the hybrid method where CFD and AI alternately calculate time series while monitoring the first principles' residuals. In our hybrid method, the role of CFD solver is not only to reduce the increased residuals but also update the network parameters with the latest CFD time series data.

The feasibility of the RePIT strategy was verified using a natural convection CFD dataset, including the heat transfer equation. First, it was observed that RePIT strategy successfully recovered the residual every time and hence the stable residual scale was maintained for 1,000 timesteps. Second, the network evolved through the transfer learning always predict the future time series in good agreement with the ground truth even in the multiple PDE condition. Third, the simulation was accelerated by 1.9 times in the assumed 2:1 ML–CFD hybrid computation condition, including the parameter-updating time. Consequently, we confirmed that our RePIT strategy is a realistic method for accelerating long-term CFD simulations. The negligible amount of training data is the reason why RePIT strategy is a universal method which can respond to geometry or boundary condition changes.

An acceleration performance comparison of the method of including all grids in the ML zone by increasing the network model/training data and the method of including the CFD solver in the boundary region is part of our future work. Investigating improved performance following network and residual tolerance optimization and observing the variation in the acceleration performance according to the CFD complexity will be the focus of future study. Additionally, the extension of this strategy to nonuniform mesh conditions with increasing complexity of the geometry remains an issue. The graphical neural networks may be a solution for nonuniform or unstructured mesh conditions. In summary, RePIT strategy is a promising technique to reduce the CFD simulation cost in industry, but more vigorous validation and optimization studies are needed.

**Nomenclature**

| | |
|---|---|
| $a_P$ | Center coefficient |
| $a_{nb}$ | Neighboring grid-point coefficient |
| $b$ | Contribution of the source term |
| $g$ | Acceleration of gravity |
| $h$ | Sum of the internal energy |
| $J(\theta)$ | Loss function |
| $K$ | Kinetic energy |
| $n$ | Number of samples |
| $n_{CFD}$ | Number of CFD timesteps |
| $n_{ML}$ | Number of ML timesteps |
| $p$ | Static pressure |
| $R^\phi$ | Residual for variable $\phi$ |
| $S$ | Source term |
| $T$ | Temperature |
| $t_{tr}$ | Network training time |
| $t_{up}$ | Parameter updating time |
| $t_{ML}$ | ML computation time for one timestep |
| $t_{CFD}$ | CFD computation time for one timestep |
| u | Velocity vector |
| $u$ | x-axis velocity |
| $v$ | y-axis velocity |
| $w$ | Weighting factor |
| $\alpha$ | Learning rate in machine learning |
| $\alpha_{eff}$ | Effective thermal diffusivity |
| $\varepsilon$ | Residual |
| $\theta$ | Parameter in machine learning |
| $\mu_{eff}$ | Effective viscosity |
| $\rho$ | Density |
| $\phi$ | General variable |

**Funding**

This research was supported by the MOTIE (Ministry of Trade, Industry, and Energy) in Korea, under the Human Resource Development Program for Industrial Innovation(Global) (P0017306, Global Human Resource Development for Innovative Design in Robot and Engineering) supervised by the Korea Institute for Advancement of Technology (KIAT) and the National Research Foundation of Korea (NRF) grant funded by the Korea government (MSIT) (RS-2023-00278121). Ricardo Vinuesa acknowledges the financial support from the Swedish Research Council (VR).

**References**

1. R. Temam, Navier-Stokes equations: theory and numerical analysis, American Mathematical Socierty 343 (2001).
2. L.I.G. Kovasznay, Laminar flow behind a two-dimensional grid, Mathematical Proceedings of

the Cambridge Philosophical Society (1948).

3. Y. Varol, A. Koca, H.F. Oztop, E. Avvi, Analysis of adaptive-network-based fuzzy inference system (ANFIS) to estimate buoyancy-induced flow field in partially heated triangular enclosures, Expert Systems with Applications 35(4) (2008) 1989-1997.

4. J. Jeon, Y.S. Kim, W. Choi, S.J. Kim, Identification of Hydrogen Flammability in steam generator compartment of OPR1000 using MELCOR and CFX codes, Nuclear Engineering and Technology 51(8) (2019) 1939-1950.

5. H. Eivazi, S.L. Clainche, S. Hoyas, R. Vinuesa, Towards extraction of orthogonal and parsimonious non-linear modes from turbulent flows, Expert Systems with Applications 202 (2022) 117038.

6. S. Ahmad, Computational analysis of comparative heat transfer enhancement in Ag-$H_2O$, $TiO_2$-$H_2O$ and Ag-$TiO_2$-$H_2O$: Finite difference scheme, Journal of the Taiwan Institute of Chemical Engineers 142 (2023) 104672.

7. S. Ahmad, T. Hayat, A. Alsaedi, Z.H. Khan, M.W.A. Khan, Finite difference analysis of time-dependent viscous nanofluid flow between parallel plates, Communications in Theoretical Physics 71 (2019) 1293.

8. S. Ahmad, Z.H. Khan, Numerical solution of micropolar fluid flow with heat transfer by finite difference method, International Journal of Modern Physics B 36 (2022) 2250037.

9. Y. Ai, X. Liu, Y. Huang, L. Yu, Numerical analysis of the influence of molten pool instability on the weld formation during the high speed fiber laser welding, International Journal of Heat and Mass Transfer 160 (2020) 120103.

10. Y. Ai, L. Yu, Y. Huang, X. Liu, The investigation of molten pool dynamic behaviors during the "∞" shaped oscillating laser welding of aluminum alloy, International Journal of Thermal Sciences 173 (2022) 107350.

11. J. Jeon and S.J. Kim, Recent progress in hydrogen flammability prediction for the safe energy systems, Energies 13(23) (2020) 6263.

12. J. Jeon, D. Shin, W. Choi, S.J. Kim, Identification of the extinction mechanism of lean limit hydrogen flames based on Lewis number effect, International Journal of Heat and Mass Transfer 174 (2021) 121288.

13. J. Jeon, Y.S. Kim, H. Jung, S.J. Kim, A mechanistic analysis of H2O and CO2 diluent effect on hydrogen flammability limit considering flame extinction mechanism, Nuclear Engineering and Technology 53(10) (2021) 3286-3297.

14. I.C. Tolias, et al., Numerical simulations of vented hydrogen deflagration in a medium-scale enclosure, Journal of Loss Prevention in the Process Industries 52 (2018) 125-139.

15. M. Morimoto, K. Fukami, K. Zhang, K. Fukagata, Generalization techniques of neural networks for fluid flow estimation, Neural Computing and Applications (2022) 1-23.

16. R. Vinuesa, et al., The role of artificial intelligence in achieving the Sustainable Development Goals, Nature Communications 11(1) (2020) 1-10.

17. J. Jeon, J. Lee, S.J. Kim, Finite volume method network for the acceleration of unsteady computational fluid dynamics: Non-reacting and reacting flows, International Journal of Energy Research 46(8) (2022) 10770-10795.

18. R. Vinuesa, R. and S.L. Brunton, Enhancing computational fluid dynamics with machine learning, Nature Computational Science 2(6) (2022) 358-366.

19. S. Lee and D. You, Data-driven prediction of unsteady flow over a circular cylinder using deep learning, Journal of Fluid Mechanics 879 (2019) 217-254.

20. X. Guo, W. Li, F. Iorio, Convolutional neural networks for steady flow approximation, 22nd ACM SIGKDD international conference on knowledge discovery and data mining (2016).

21. M. Raissi, P. Perdikaris, G.E. Karniadakis, Physics-informed neural networks: A deep learning framework for solving forward and inverse problems involving nonlinear partial differential equations, Journal of Computational physics 378 (2019) 686-707.

22. M.S. Go, J.H. Lim, S. Lee, Physics-informed neural network-based surrogate model for a virtual thermal sensor with real-time simulation, International Journal of Heat and Mass Transfer 214 (2023) 124392.

23. L. Lu, X. Meng, Z. Mao, G.E. Karniadakis, DeepXDE: A deep learning library for solving differential equations, SIAM review 63(1) (2021) 208-228.

24. L. Lu, P. Jin, G. Pang, Z. Zhang, G.E. Karniadakis, Learning nonlinear operators via DeepONet based on the universal approximation theorem of operators, Nature Machine Intelligence 3(3) (2021) 218-229.

25. Z. Li, et al., Fourier neural operator for parametric partial differential equations. arXiv preprint, arXiv:2010.08895, 2020.

26. H. Eivazi, L. Guastoni, P. Schlatter, H. Azizpour, R. Vinuesa, Recurrent neural networks and Koopman-based frameworks for temporal predictions in a low-order model of turbulence, International Journal of Heat and Fluid Flow 90 (2021) 108816.

27. P.A. Srinivasan, L. Guastoni, H. Azizpour, P. Schlatter, R. Vinuesa, Predictions of turbulent shear flows using deep neural networks, Physical Review Fluids 4(5) (2019) 054603.

28. B. Stevens and T. Colonius, FiniteNet: A fully convolutional LSTM network architecture for time-dependent partial differential equations, arXiv:2002.03014 (2020).

29. A. Takbiri-Borujeni and M. Ayoobi. Application of physics-based machine learning in combustion modeling, 11th US National Combustion Meeting (2019).

30. C.J. Greenshields, Openfoam user guide version 6, The OpenFOAM Foundation 237 (2018) 624.

31. A. Ansari, S. Mohaghegh, M. Shahnam, J.F. Dietiker, T. Li, Data driven smart proxy for cfd application of big data analytics & machine learning in computational fluid dynamics, report two: Model building at the cell level, National Energy Technology Laboratory (NETL), Pittsburgh, PA, Morgantown, 2018.

32. L. Guastoni, et al., Convolutional-network models to predict wall-bounded turbulence from wall quantities, Journal of Fluid Mechanics 928 (2021) A27.

33. A. Krizhevsky, I. Sutskever, G.E. Hinton, Imagenet classification with deep convolutional neural networks, Advances in Neural Information Processing Systems 25 (2012).

34. D.E. Rumelhart, G.E. Hinton, R.J. Williams, Learning representations by back-propagating

errors, Nature 323(6088) (1986) 533-536.

35. S. Kim, W. Ji, S. Deng, Y. Ma, C. Rackauckas, Stiff neural ordinary differential equations, Chaos: An Interdisciplinary Journal of Nonlinear Science 31(9) (2021).

36. H. Hajibeygi and P. Jenny, Adaptive iterative multiscale finite volume method. Journal of Computational Physics 230 (2011) 628-643.

37. ANSYS FLUENT 18.0 Theory Guide, 2017.

38. R. Kumar and A. Dewan, URANS computations with buoyancy corrected turbulence models for turbulent thermal plume, International Journal of Heat and Mass Transfer 72 (2014) 680-689.

39. L.W. Kit, H. Mohamed, N.Y. Luon, L. Chan, Numerical Simulation of Ventilation in a Confined Space, Journal of Advanced Research in Fluid Mechanics and Thermal Sciences 107 (2023) 1-18.

40. P.V. Nielsen, Flow in air conditioned rooms, Ph. D. thesis from the Technical University of Denmark, 1976.

41. N. Srivastava, G. Hinton, A. Krizhevsky, I. Sutskever, R. Salakhutdinov, Dropout: a simple way to prevent neural networks from overfitting, The Journal of Machine Learning Research 15(1) (2014) 1929-1958.

42. D.P. Kingma and J. Ba, Adam: A method for stochastic optimization, arXiv:1412.6980 (2014).

43. J. Jeon, J. Lee, H. Eivazi, R. Vinuesa, S.J. Kim, Physics-Informed Transfer Learning Strategy to Accelerate Unsteady Fluid Flow Simulations, arXiv:2206.06817 (2022).